\documentclass[flushrt,preprint]{aastex}       % onecolumn (second format)
\usepackage{graphicx}
\usepackage{amsmath,amssymb}
\usepackage[hyphens]{url}
\usepackage{natbib}
\usepackage{caption,subcaption}
\begin{document}
\sloppy

\title{Decreasing World Aridity in a Warming Climate}
\shorttitle{Decreasing World Aridity in a Warming Climate}
\shortauthors{Wang and Katz}
\author{S. Wang}
\affil{Department of Physics}
\affil{Washington University, St. Louis, Mo. 63130}
\author{J.~I.~Katz}
\affil{Department of Physics and McDonnell Center for the Space Sciences}
\affil{Washington University, St. Louis, Mo. 63130}
\affil{Tel.: 314-935-6276; Facs: 314-935-6219}
\email{katz@wuphys.wustl.edu}
\begin{abstract}
The mean world climate has warmed since the 19th Century as the
anthropogenic emission of greenhouse gases has increased the atmospheric
opacity to thermal infrared radiation.  Has this warming increased the
frequency or severity of droughts?  We define an objective aridity index
that quantifies the precipitation forcing function of drought.  Using the
GHCN daily database from {\it c.\/}~1900 to the present and averaging over
hundreds or thousands of sites in each of eight continental or
semi-continental regions, we evaluate trends in aridity.  Seven of these
regions have sufficient data for significant conclusions.  In them the mean
aridity decreased at logarithmic rates from $(-2.2 \pm 0.6) \times 10^{-4}$
to $(-5.1 \pm 0.2) \times 10^{-3}$ per year, three or more times their
nominal ($1\sigma$) uncertainties.
\end{abstract}
\maketitle
\clearpage
\section{Introduction}
\label{sec:intro}
How do climatic variables other than temperature change under conditions of
anthropogenic global warming?  Although it is difficult to predict the
future climate in quantitative detail, the past may be a guide.  Historic
climatological databases are extensive, and it is possible to study changes
over the approximately 150 years since the beginning of anthropogenic
warming.  Studies in this series have addressed trends in storminess
\citep{MK13,CKK18}, temperature extremes \citep{FK17,FK18}, pressure
variance \citep{HK19} and aridity \cite{FCKK16}.

There are many definitions of drought \citep{KD02}.  Drought is not
determined by precipitation alone, because temperature, insolation, humidity
and wind affect rates of evaporation and evapotranspiration, precipitation
is the ultimate forcing function and is readily quantified, with extensive
historic databases.  A dearth of precipitation over a few months of a
growing season may be drought to the farmer, even if the annual total
precipitation is high.  Our approach is entirely empirical, defining an
aridity index that describes precipitation on the seasonal time scales of
most importance to agriculture and calculating its trends over historic
time.

In earlier work \citep{FCKK16} we studied precipitation trends in the 48
contiguous United States during the era of anthropogenic warming.  Here we
extend that study to trends on other continents.  The same methods and
definitions are used, so much of the Introduction and Methods sections here
follow those sections of the earlier paper.  Because of the long term
persistence of annual precipitation and other meteorological and
climatological time series on historic (century or longer) time scales
\citep{MW69,P98,MZ14,MKC19}, it may be difficult to attribute trends to
anthropogenic forcing, as opposed to persistent intrinsic variation, but the
first step is to measure those trends.  Comparison of trends at different
sites in different climate regimes also permits testing the ``DDWW''
hypothesis \citep{T11} that while warming increases the rate of ocean
evaporation and hence the world-wide total precipitation, dry regions get
drier while wet regions get even wetter.  That is a significant practical
issue because water resources are of greatest concern in dry regions.

%Longer term studies \citep{CWEMS04,D11b} suggest a
%correlation between proxy drought measurements in North America and the
%(northern European) Medieval Warm Period.  Their applicability to the modern
%period of warming by greenhouse gases is uncertain.

The widely used Palmer Drought Index (PDI) and the Palmer Drought Severity
Index (PDSI) \citep{P65,K86,D11a,D11b} involve a complex interplay among
precipitation and modeled evapotranspiration, soil moisture, runoff and
recharge, but do not include the effects of atmospheric humidity,
vegetation, cloud cover, precipitation rate, wind and soil permeability that
are difficult to measure except on a micro-scale.  These indices filter the
precipitation forcing through a complex and model-dependent transfer
function.  They are useful to agriculturalists and water resource and flood
control engineers because they measure the deviation of local conditions
from their long-term means that are the basis of farming and planning.
However, we wish to isolate precipitation forcing from other processes,
including temperature change, that contribute to the response of the
hydrological system.

Studies of drought trends using the PDSI \citep{KH90,DTK98,DTQ04,D11a,T14}
are inconsistent, with the earliest work finding no evidence of a trend but
some more recent work \citep{D11b,D13} indicating a drying trend during a
period that mostly overlaps with that considered here.  Because the PDSI
includes soil drying as a result of increased evaporation as the climate
warms, it conflates effects of precipitation and temperature changes
\citep{SWR12} and is not a direct measure of the precipitation forcing
function.  Warming increases evaporation and evapotranspiration and the
climatological drought described by the PDSI. 

A number of other drought indices exist \citep{Q09,D11b}, but are also
imperfect tools for studying the possible effects of climate change on
precipitation.  For example, the Standardized Precipitation Index (SPI)
\citep{MK93,MK95,G98,G99} compares the precipitation at a site over some
period (chosen in the range from one month to several years) to the
statistical distribution of recorded precipitation at that site in periods
of the same length.  This identifies anomalous (unusually dry or wet)
periods that are then assigned an index value based on the fraction of
such periods in the record that were drier or wetter.  The SPI describes
how unusual is the rate of precipitation at that site in comparison to its
history, without making the unproven assumption of a Gaussian or any other
distribution, but does not quanitfy the magnitude of its deviation from the
mean or its implications.
%Studies using the SPI may be more directly comparable to ours, but are few.
%For example, \cite{GR03} modeled drought in the northeastern United States
%and \cite{LBS10} analysed SPI data for Hungary, but neither of these
%publications present detail sufficient for comparison.

Dearths of precipitation are basic and elemental climate parameters.  We
separate changes in precipitation from those of temperature with which they
are conflated in many drought indices:  Has the frequency and severity of
periods of low precipitation changed?  This question can be addressed by
comparatively simple and unambiguous statistical measures, free of the
complications inherent in these drought indices of including parameters,
such as humidity, vegetation, cloud cover, precipitation rates and wind, for
which data are absent or limited, but that affect evaporation,
evapotranspiration and runoff.

The increase of water vapor pressure with increasing temperature given by
the Clausius-Clapeyron equation implies that warming is accompanied by
increasing ocean evaporation and therefore increasing globally-averaged
precipitation.  However, this does not address questions of its spatial and
temporal distributions, and hence of drought.

The purpose of this work is to determine if extended periods of low
precipitation that are the forcing function of drought have become more, or
less, frequent or severe around the world as the climate has warmed in the
last approximately 150 years.  A few studies of drought trends over periods
of a century or more, long enough to cover most of the period of
anthropogenic warming, exist, but have generally been limited to
comparatively few sites and restricted geographic areas (for example,
\cite{QZ01,LZQ02,QYZ06,MKC19}).
\section{Methods}
\label{sec:methods}
In order to avoid the uncertainties of modeling \citep{D13}, our approach is
entirely empirical.  We forgo any attempt to interpret our historical
results as tests of the validity of climate models or of their predictions
of drought.  Because of the ``red'' spectrum of natural climatic variation
\citep{MW69,P98,MZ14,MKC19} and its complex dependence on space, time and
the variable considered, we do not attempt the difficult task of separating
natural variations on time scales of 100 years or more from anthropogenic
climate change.

We define an empirical aridity index that measures any seasonal dearth
of precipitation, compute its averages over intervals of 11 years, and
determine or bound any long term trends in the interval averages over the
period 1898--2018.  We choose 11 year, rather than decadal, intervals, so
that any possible climatic effects of the 11 year Solar cycle
\citep{HS97,MAMSvL09,NAS12} are not aliased into a spurious slow trend or
long period.  Our aridity index is a reciprocal function of the
precipitation in three month periods because dry seasons are more
significant for agriculture than the total annual (or longer term) rainfall.
It is unrelated to aridity indices \citep{Wiki15} that use long-term
averaged precipitation and evapotranspiration to describe mean climate.  Our
index captures only dearths of precipitation.  It does not address dearths
of snowpack that may result from warming temperatures rather than from
dearths of precipitation.

We use the Global Historical Climate Network Daily database \citep{GHCN}.
For each site we calculate precipitation totals $P_{i,j}$ for the
three-month periods approximately corresponding to the seasons
January--March, April--June, July--September and October--December, where
$i$ denotes the site and $j$ the calendar quarter and year.  We define an
annual mean seasonal aridity index at the $i$-th site:
\begin{equation}
\label{indexdefine}
A_{i,Y} \equiv \frac{1}{4} \sum\limits_{j \in Y} \frac{P_0^2}
{(P_{i,j} + P_0)^2},
\end{equation}
where $Y$ denotes the year.  $P_0$ regularizes and normalizes the index,
keeping it a smooth and non-singular function of the precipitation.  If
there is no precipitation at all $A_{i,Y} = 1$, while if there is any
precipitation $0 < A_{i,Y} < 1$.  $A_{i,Y}$ is insensitive to variations in
precipitation that are either $\ll P_0$ or $\gg P_0$.  If $P_0$ approximates
the quarterly precipitation requirement of important crops then $A_{i,Y}$ is
most sensitive to values of $P_{i,j}$ that are near the threshold of
sufficiency for those crops, a desirable property.

This index is influenced by the severity of dry periods but, like the
farmer, is not much affected by the difference between zero precipitation
and agriculturally insignificant amounts $\ll P_0$.  It is little affected
by variations in precipitation in wet periods; enough is enough.  It
embodies temporal information not contained in annual precipitation values
\citep{D11b,D13}; the agriculturalist is chiefly concerned with dry seasons,
which are hardly mitigated by intervening wet seasons.

The value of $P_0$ is necessarily arbitrary.  We take $P_0 = 6\, \mathrm{in}
= 15.24$\,cm.  This choice is about one quarter of the mean annual
precipitation that distinguishes rain-fed wet from dry land agriculture.
Agriculturally insignificant precipitation ($P_{i,j} \ll 6\, \mathrm{in}$)
has little effect on a quarter's contribution to $A_{i,Y}$.  For $P_{i,j} >
12\,\mathrm{in}$, ample rain, $A_{i,Y} < 0.11$.  In order to avoid bias
resulting from the omission of a season (that might be seasonally dry or
wet) when only incomplete data are available, $A_{i,Y}$ is not computed if
precipitation data are not available for the entire year, and that year is
excluded from the analysis.

The mean seasonal aridity index in the interval $I$
\begin{equation}
\langle A_i \rangle_I \equiv \frac{1}{N_{i,I}} \sum\limits_{Y \in I}
A_{i,Y},
\end{equation} 
where $N_{i,I}$ is the number of years with complete data in the $I$-th
interval (in the period 1898--2018 $I = 1,2,\dots 11$).  If fewer than six
years have complete data $\langle A_i \rangle_I$ is not computed and data
for that site and 11 year interval are ignored.  The
uncertainty of $\langle A_i \rangle_I$ is estimated:
\begin{equation}
\sigma_{i,I} \equiv \frac{1}{N_{i,I}} \sqrt{\sum\limits_{Y \in I}
\left(A_{i,Y} - \langle A_i \rangle_I\right)^2}.
\end{equation}

For each site $i$ we calculate the mean aridity index
\begin{equation}
\langle A_i \rangle \equiv \frac{1}{{\cal N}_i} \sum\limits_I \langle A_i
\rangle_I,
\end{equation}
where ${\cal N}_i$ is the number of intervals $I$ with valid data.  The time
dependence of the $\langle A_i \rangle_I$ is described by a least-squares
linear fit to the logarithmic slope per year $S_i$:
\begin{equation}
\langle A_i \rangle_I = 11 I \langle A_i\rangle \times S_i + C_i.
\label{logslope}
\end{equation}
We use logarithmic slopes so that very dry sites, with large aridity and
possibly large slopes of aridity, are not overweighted in the averages.  The
aridity ranges over about an order of magnitude for our sites, so that if
the slopes themselves were used their means would be dominated by the most
arid sites.  Positive values of $S_i$ indicate increasing aridity, and {\it
vice versa\/}.

We only consider sites with a minimum of four 11-year intervals with valid
(at least six years with data) $\langle A_i \rangle_I$.  Of 108081 sites in
the database, only 15068 satisfy this criterion.  In eight of our regions
the mean number of valid intervals of sites we use (those with a minimum of
four valid intervals) is between five and six, while in South America it is
4.79.  If more valid intervals were required, the baseline over which trends
were averaged would be lengthened but the number of sites over which we
could compute averages would be fewer, and {\it vice versa\/}.  The tradeoff
is unavoidable and any chosen criterion for inclusion is necessarily
arbitrary.

%The logarithmic slope is defined
%\begin{equation}
%\label{logslope}
%{\cal L}_i \equiv {S_i \over \langle \langle A_i \rangle_I \rangle}.
%\end{equation}
%Use of ${\cal L}_i$ rather than $S_i$ in averages over sites with widely
%differing aridities avoids giving sites with large aridity, and therefore
%typically large slopes and random fluctuations, disproportionate influence.
\section{Results and Discussion}
\label{sec:results}
We divide the world into nine regions, as shown in Fig.~1.  These regions
are either continents or semi-continents.  The definitions of the
semi-continents roughly correspond to climatic regions.  Africa is divided
in the Sahel into humid Subsaharan Africa and arid North Africa (including
the arid Middle East with North Africa).  Asia is divided along the latitude
of the Himalayas into tropical South Asia and temperate and subarctic North
Asia.  The results are summarized in Table~1 that defines the regions and
gives the mean logarithmic slope of the aridity in each.

\begin{figure}[t!]
	\centering
	\includegraphics[width=\columnwidth]{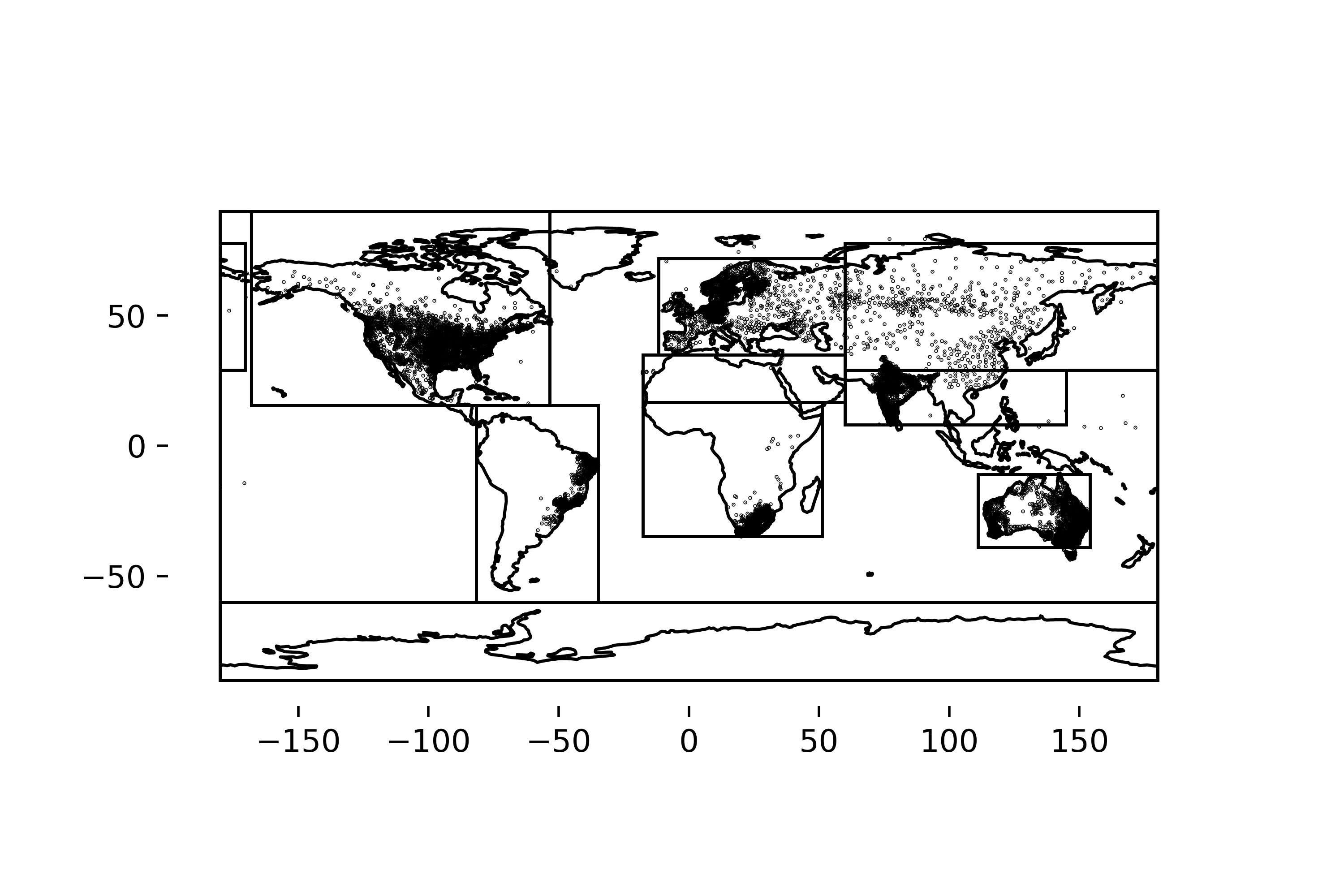}
	\caption{\label{regionsmap}Locations of sites with sufficient data
	for inclusion.  The world is divided into the same eight continental
	or semi-continental regions as by \cite{FK18}, but the sites differ
	because of differences in the databases and variables studied.
	Sites not in any of the eight regions, mostly on ocean islands, are
	collected as Undefined.}
\end{figure}

\begin{table}[h!]
	\begin{center}
		\begin{tabular}{|l|cc|r|c|c|}
			\hline
			Region & Latitudes & Longitudes & $N$ \ \quad& Slope (/y) & $r$ \\
			\hline
			Australia & 10.93--39.1 S & 111--154 E & 3881 & $-0.41 \pm 0.03 \times 10^{-3}$ & $+0.04$ \\
			Europe & 35--72 N & 11.55 W--60 E & 2463 & $-2.49 \pm 0.07 \times 10^{-3}$ & $-0.12$ \\
			North Africa \& ME & 16.7--35 N & 17.7 W--60 E & 17 & $+0.58 \pm 0.34 \times 10^{-3}$ & $+0.04$ \\
			North America & 15.5--90 N & 53.5--168 W & 3897 & $-0.97 \pm 0.04 \times 10^{-3}$ & $+0.08$ \\
			North Asia & 29--77.7 N & 60 E--170.4 W & 670 & $-1.50 \pm 0.10 \times 10^{-3}$ & $+0.18$ \\
			South America & 15.5 N--60 S & 34.9--81.7 W & 1000 & $-5.15 \pm 0.19 \times 10^{-3}$ & $+0.42$ \\
			South Asia & 8.15--29 N & 60--145 E & 1774 & $-0.22 \pm 0.06 \times 10^{-3}$ & $-0.10$ \\
			Sub-Saharan Africa & 34.8 S--16.7 N & 17.7 W--51.2 E & 1160 & $-0.33 \pm 0.06 \times 10^{-3}$ & $-0.08$ \\
			Undefined & & & 206 & $+0.13 \pm 0.24 \times 10^{-3}$ & $-0.16$ \\
			\hline
		\end{tabular}
	\end{center}
	\caption{\label{results}Mean logarithmic derivatives of the aridity
	(slopes) and its Pearson correlation $r$ with the mean aridity in
	nine regions.  The slopes for North Africa \& Middle East and
	``Undefined'' (every site, mostly ocean islands, not in one of the
	defined continental regions), with comparatively few sites, are not
	significantly different from zero.  In each of the remaining seven
	regions the aridity decreased, apparently significantly.  The errors
	indicated are the nominal $1\sigma$ uncertainties of the mean
	slopes, defined as the standard deviations of the individual site
	slopes in the region, divided by $N^{1/2}$, where $N$ is the number
	of sites in the region with sufficient valid data for inclusion.
	This can overestimate the uncertainty of the mean, if climate
	evolves differently at different sites, and underestimate it by
	ignoring correlations between sites.  In every region except South
	America $r$ is small, inconsistent with the DDWW hypothesis.}
\end{table}

The mean aridities and their natural logarithmic slopes (time derivative
divided by the mean $\langle A_i \rangle$; Eq.~\ref{logslope}) at each
site (with the exception of a few sites that lie outside the bounds of the
graphs) in each of these regions are shown in Fig.~2.  The Pearson
correlations in Table 1 and the scatter plots in Fig.~2 indicate that,
except for South America, there is very little correlation between aridity
and its trend; the DDWW hypothesis is not confirmed.  This agrees with a
conclusion \citep{Greve14,G18} based on different datasets over a shorter
(1948--2005) interval as well as our earlier results for the 48 contiguous
United States \citep{FK18}.

\begin{figure}
	\centering
	\begin{subfigure}{0.47\textwidth}
		\centering
		\includegraphics[width=0.99\linewidth,height=4.5cm]{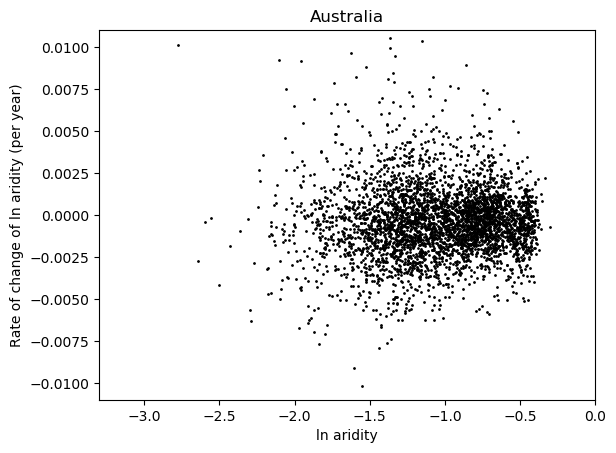}
	\end{subfigure}
	~
	\begin{subfigure}{0.47\textwidth}
		\centering
		\includegraphics[width=0.99\linewidth,height=4.5cm]{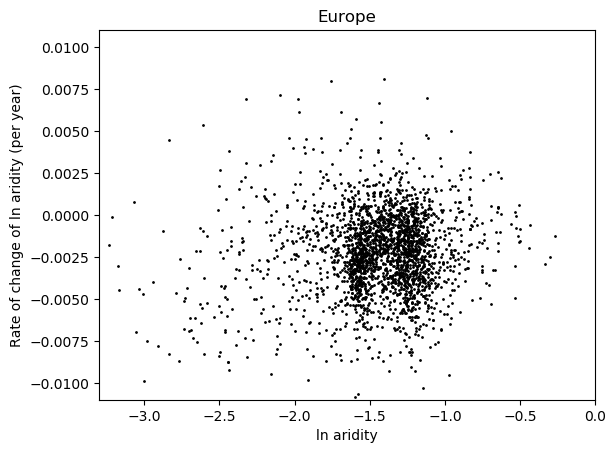}
	\end{subfigure}
	
%	\begin{subfigure}{0.47\textwidth}
%		\centering
%		\includegraphics[width=0.99\linewidth,height=4.5cm]{NorthAfricaAndME_new.png}
%	\end{subfigure}
%        ~
	\begin{subfigure}{0.47\textwidth}
		\centering
		\includegraphics[width=0.99\linewidth,height=4.5cm]{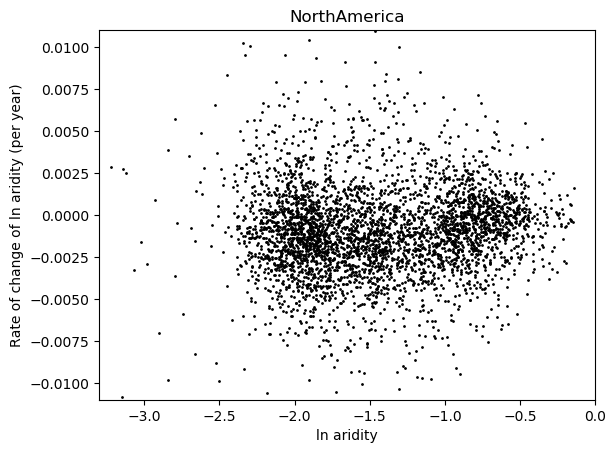}
	\end{subfigure}
	~
	\begin{subfigure}{0.47\textwidth}
		\centering
		\includegraphics[width=0.99\linewidth,height=4.5cm]{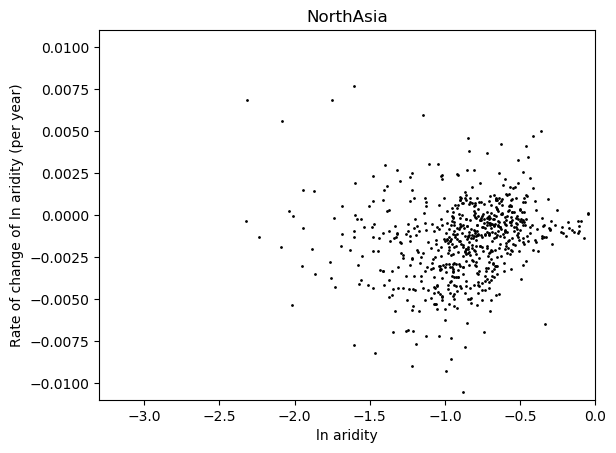}
	\end{subfigure}
	
	\begin{subfigure}{0.47\textwidth}
		\centering
		\includegraphics[width=0.99\linewidth,height=4.5cm]{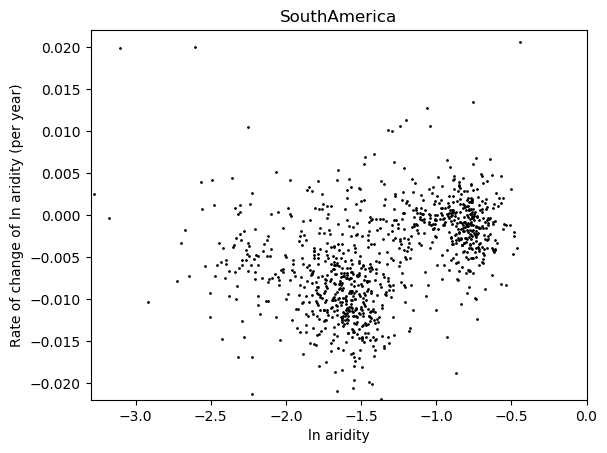}
	\end{subfigure}
	~
	\begin{subfigure}{0.47\textwidth}
		\centering
		\includegraphics[width=0.99\linewidth,height=4.5cm]{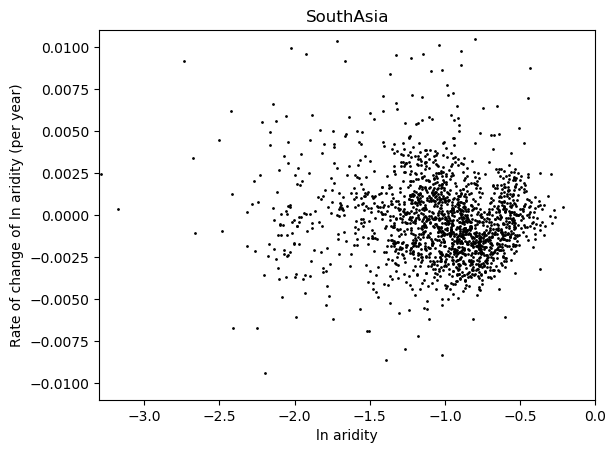}
	\end{subfigure}
	
	\begin{subfigure}{0.47\textwidth}
		\centering
		\includegraphics[width=0.99\linewidth,height=4.5cm]{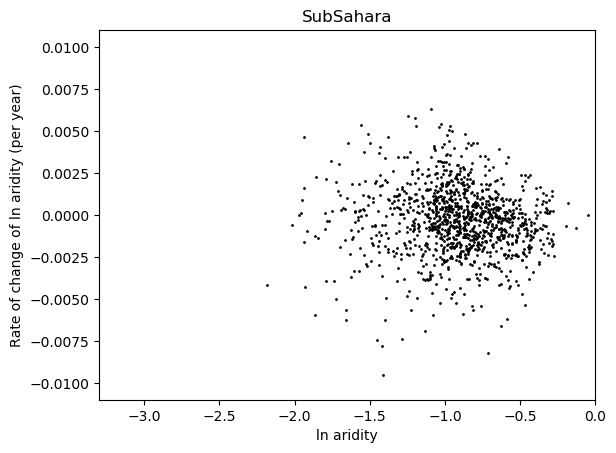}
	\end{subfigure}
	~
	\begin{subfigure}{0.47\textwidth}
		\centering
		\includegraphics[width=0.99\linewidth,height=4.5cm]{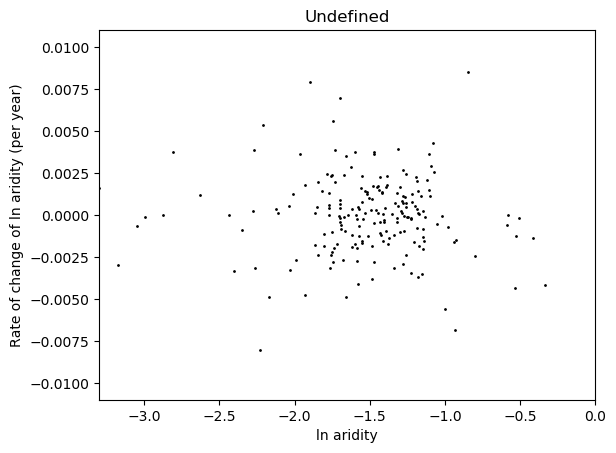}
	\end{subfigure}
	\caption{\label{distributions}Logarithmic rates of change of aridity
	index per year {\it vs.\/} $\ln{\text{mean aridity}}$ for sites in
	eight of the geographic regions defined in Fig.~\ref{regionsmap}
	(Undefined includes all sites, mostly on ocean islands, not in one
	of the other eight regions).  North Africa and Middle East has only
	17 sites with long enough data series to meet our criteria, and is
	not shown.  At a few sites the aridity changed too rapidly to appear
	within the bounds of the graphs.  The range in ordinates of the
	South America plot is twice those of the other plots to accomodate
	the rapidly decreasing aridity of its wet sites; this appears to be
	a regional phenomenon.} 
\end{figure}

The South American data, indicating a significant correlation between
aridity and its trend, appear to support the DDWW hypothesis.  However,
Figs.~1 and 3 show that nearly all the utilized (because they extend over a
minimum of four valid 11-year intervals) South American data come
from two clusters in coastal Brazil, one in and around the states of S\~{a}o
Paulo and Rio de Janeiro and one in the Northeast; of 6523 South American
sites in the database only 1000 meet our criterion for inclusion.  The
distribution of South American aridities in Fig.~2 is bimodal, dividing
around $\ln{\langle A \rangle} = -1.35$.  Fig.~3 shows that the sites in the
S\~{a}o Paulo and Rio de Janeiro areas have $\ln{\langle A \rangle} < -1.35$
while nearly all the sites in the Northeast (with the exception of a few on
its southern coast) have $\ln{\langle A \rangle} > -1.35$.  The bimodality
is an artefact of the distribution of sites with data, and the correlation
of trend with aridity is an artefact of the concentration of data in two 
regions with distinct climates and trends; despite the large number of
sites, there are effectively only two independent climatic regimes sampled.

\begin{figure}
	\centering
	\includegraphics[width=0.99\columnwidth]{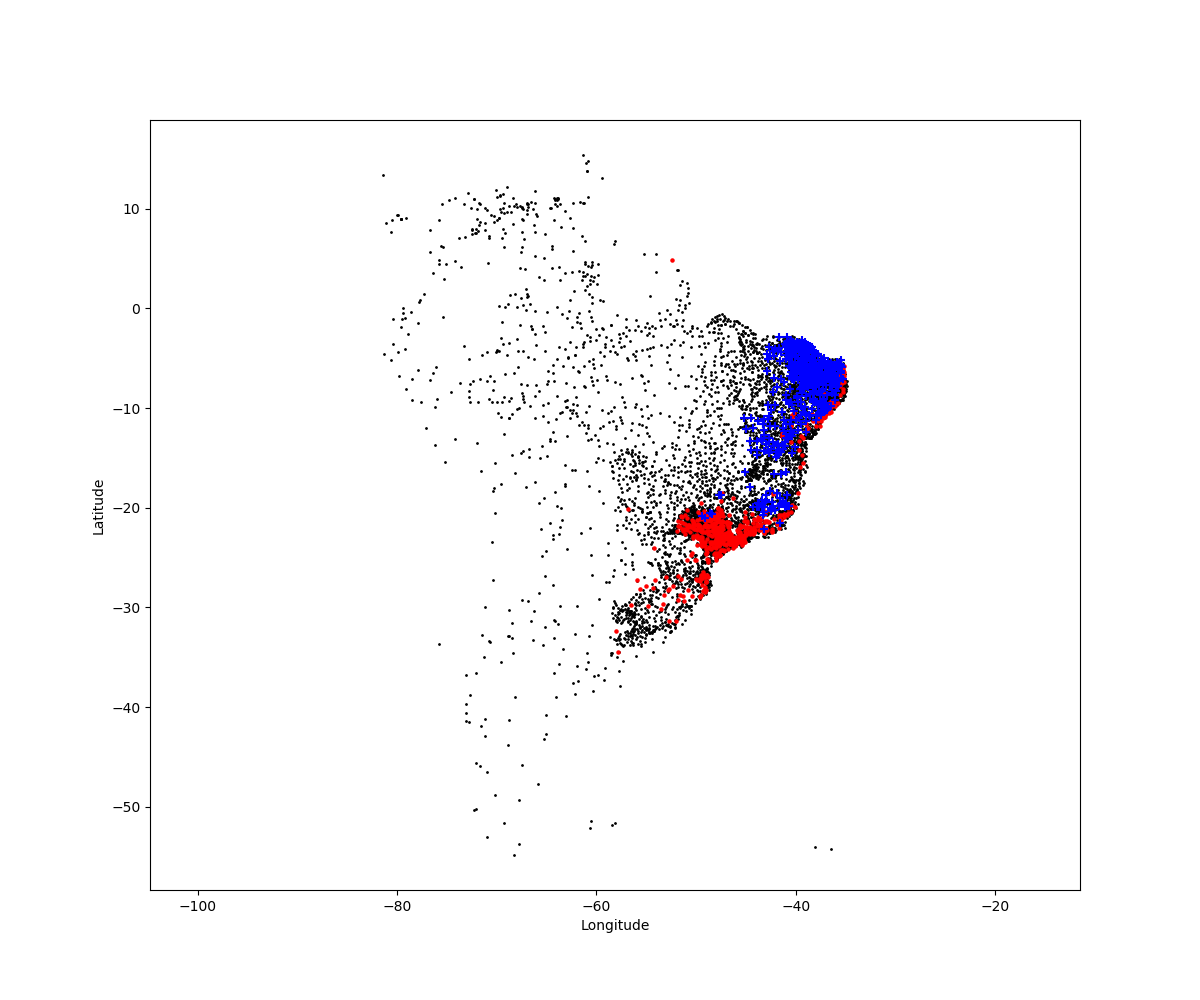}
	\caption{\label{SAsites}South American sites.  Wetter sites
	($\ln{\langle A \rangle} < -1.35$) with sufficient data by our
	criteria to establish trends are concentrated in the S\~{a}o Paulo
	and Rio de Janeiro areas and are shown as red; drier sites 
	($\ln{\langle A \rangle} > -1.35$) with sufficient data are
	concentrated in Northeast Brazil and are shown as blue.  Several
	thousand other South American sites have insufficient data to
	establish trends (and hence not shown in Fig.~2) are black dots.
	The fact that the wetter region is getting wetter while there is
	very little trend in the drier region may be specific to these two
	distinct climate regimes and not a general indication of a DDWW
	correlation, which is not found elsewhere in our data.}
\end{figure}

\begin{figure}
	\centering
	\includegraphics[width=0.5\columnwidth]{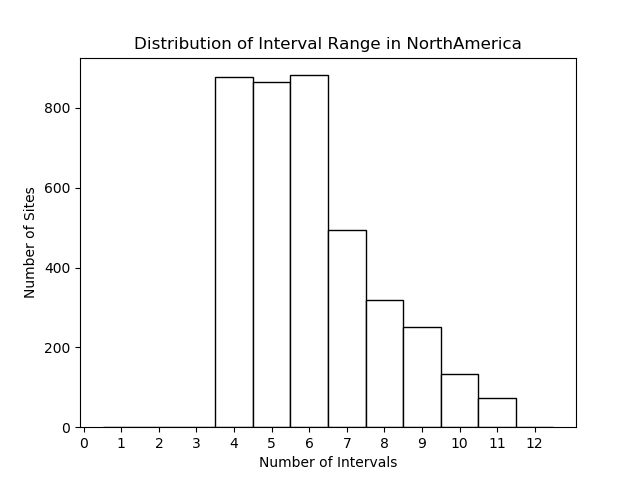}
	\caption{\label{histo}Distribution of span of 11-year intervals
	with valid data for North American sites.  Only sites with at least
	four such intervals are used.  The span between the midpoints of the
	earliest and latest valid intervals is at least 11 years $\times$
	one less than the number of valid intervals (110 years at some
	sites), but is greater if some intermediate intervals lack
	sufficient valid data for inclusion.  Other regions generally lack
	such long spans \citep{GHCN}.}
\end{figure}

Each site may have different starting and ending dates for the data, as well
as different dates of missing data between those limits.  In Fig.~4 we show
the distribution of data durations for our North American sites.

%OMITTED BECAUSE DATA NOW INCLUDED IN SCATTERPLOT
%\begin{figure}
%\begin{center}
%\includegraphics[width=4.5in]{weatherdata.d/stratified.eps}
%\end{center}
%\caption{\label{stratified}Mean slope of log aridity (per year) in each
%hexile of mean aridity distribution.  The two points with largest aridity
%describe the 50 and 18 driest sites, and are subsets of the sixth hexile.
%Error bars are one standard deviation of the means of each hexile or bin.
%All mean slopes are negative and, except for the driest hexile and its
%subsets, significantly different from zero at the level of three standard
%deviations.  Evidence for a dependence of mean log slope on aridity is only
%suggestive.}
%\end{figure}

Mean precipitation is expected to increase with global warming because of
the increase in water saturation vapor pressure with increasing temperature.
A warmer atmosphere can hold more water vapor
\citep{PAS07,BMH13,ALZLKB14,BHTPSC14}, decreasing aridity if that vapor
precipitates.  The Clausius-Clapeyron relation indicates $d\ln{p_{vap}}/dT
\approx 0.07/ ^{\circ}$K at typical temperate zone near-surface
temperatures.  The mean observed warming rate $dT/dt \approx 6 \times
10^{-3\,\circ}$K/y ($\approx 1^{\,\circ}$K over the last 160 years) then
implies, assuming constant relative humidity, water vapor content of the
atmosphere increasing at the rate
\begin{equation}
{d\ln{p_{vap}} \over dt} = {d\ln{p_{vap}} \over dT} {dT \over dt} \approx
4 \times 10^{-4}/\mathrm{y}.
\label{pvap}
\end{equation}

If all $P_{i,j} = P_q$ for a quarterly precipitation $P_q \propto p_{vap}$,
where $P_q$ is a weighted average over all sites and quarters, then,
differentiating Eq.~\ref{indexdefine} and using Eqs.~\ref{logslope} and
\ref{pvap}, the Clausius-Clapeyron relation predicts a logarithmic trend
\begin{equation}
S = - 2 {P_q \over P_q + P_0} {d\ln{p_{vap}} \over dt} \approx
	- {P_q \over P_q + P_0}\ 0.8/\mathrm{millennium},
\end{equation}
For a representative $P_q = 8\,\mathrm{in}$, $S \approx - 0.45/
\mathrm{millennium}$.  This is qualitatively consistent with the results of
Table~1, but does not explain the large differences among continents.
\section{Conclusions}
Using a worldwide database that extends over almost the entire period of
significant anthropogenic emission of greenhouse gases, we have found, in
every region for which we have enough data for significant results, a
statistically significant mean decreasing trend aridity.  Our metric is
chosen to reflect agriculturally significant drought, and is insensitive to
annual mean rainfall or changes in rainfall in wetter or very dry regions.

The logarithmic rates of decrease of aridity, although each (with the
exceptions of regions with few data) nominally very significant, also
differ by many times their nominal uncertainties.  This climatic trend is
not uniform around the world, in agreement with a result we found previously
\citep{FK18} on a regional scale for the 48 contiguous United States.  

Our results disagree with the hypothesis that drier regions are getting
drier.  In all regions except South America there is no evident correlation
between aridity and its rate of change.  In South America the wetter sites
are rapidly increasing in wetness, but the drier sites are not, on average,
getting significantly drier, although there is substantial apparent random
scatter.

The mean trend is qualitatively consistent with expectations for the mean
precipitation from the increase of water saturation vapor pressure as the
climate has warmed, but our aridity index distinguishes periods of drought
from the mean precipitation.  We have found that the previously predicted
increase in mean precipitation is also reflected in a reduction, within the
area of our data, in periods of drought.  This result (it is conceivable
that increasing mean precipitation might coexist with more severe local or
regional droughts even though mean global precipitation increases) may
inform discussions of the nature and consequences of
climate change. 

\clearpage
\bibliography{droughtworldwang}   % name your BibTeX data base

%\begin{thebibliography}{}
% etc
%\end{thebibliography}

\end{document}